# A MORE APPROPRIATE PROTEIN CLASSIFICATION USING DATA MINING

[1]MUHAMMAD MAHBUBUR RAHMAN, [2]ARIF UL ALAM, [3]ABDULLAH-AL-MAMUN, [4]TAMNUN E MURSALIN

[1] Lecturer, Department of CSE, American International University-Bangladesh,

[2] Lecturer, Department of ETE, University of Liberal Arts Bangladesh,

[4]Assoc Prof., Department of CSE, University of Liberal Arts Bangladesh

**ABSTRACT**

Research in bioinformatics is a complex phenomenon as it overlaps two knowledge domains, namely, biological and computer sciences. This paper has tried to introduce an efficient data mining approach for classifying proteins into some useful groups by representing them in hierarchy tree structure. There are several techniques used to classify proteins but most of them had few drawbacks on their grouping. Among them the most efficient grouping technique is used by PSIMAP. Even though PSIMAP (Protein Structural Interactome Map) technique was successful to incorporate most of the protein but it fails to classify the scale free property proteins. Our technique overcomes this drawback and successfully maps all the protein in different groups, including the scale free property proteins failed to group by PSIMAP. Our approach selects the six major attributes of protein: a) Structure comparison b) Sequence Comparison c) Connectivity d) Cluster Index e) Interactivity f) Taxonomic to group the protein from the databank by generating a hierarchal tree structure. The proposed approach calculates the degree (probability) of similarity of each protein newly entered in the system against of existing proteins in the system by using probability theorem on each six properties of proteins. This function generates probabilistic value for deriving its respective weight against that particular property. All probabilistic values generated by six individual functions will be added together to calculate the bond factor. Bond Factor defines how strongly one protein bonds with another protein base on their similarity on six attributes. Finally, in order to group them in hierarchy tree, the aggregated probabilistic value will be compared with the probabilistic value of the protein that resides at the root. If there is no root protein (i.e. at the initial state), the first protein will be considered as the root and depending on the probabilistic value it can change its relative position. Recursively, at each node, we have applied this technique to calculate the highest probable position for a particular protein in the tree.

**Keywords:** *Bioinformatics, Protein, Protein Grouping Techniques, PSIMAP, Scale Free Protein.*

## 1. INTRODUCTION:

Classification of protein based on their various properties is a crucial issue in different fields of biological science. Researches in pharmacy, biochemistry, genetic engineering even in agriculture vastly rely on appropriate protein grouping techniques. Emphasizing the importance of protein classification some research groups in bioinformatics have initiated their projects with a view to deriving appropriate algorithms for protein classification. Protein can be classified based on their some properties, namely, a) Structure comparison b)Sequence Comparison c) Connectivity d) Cluster Index e) Interactivity f) Taxonomic and age diversity**[1].** Individual research group, so far has attempted to classify protein focusing on only one or two above stated properties. As for example, BMC bioinformatics research group has developed an in silico classification system entitled HODOCO (Homology modeling, Docking and Classification Oracle), in which protein Residue Potential Interaction Profiles (RPIPS) are used to summarize protein -protein interaction characteristics. This system applied to a dataset of 64 proteins of the death domain super family this was used to classify each member into its proper subfamily. Two classification methods were attempted, heuristic and support vector machine learning. Both methods were tested with a 5-fold cross-validation. The heuristic approach yielded a 61% average accuracy, while the machine learning approach yielded an 89% average accuracy. Though this is a good technique but it concentrates on only protein-protein interaction property **[2].**

Wan K. Kim, Dan M. Bolser and Jong H. Park **[1]** had used PSIMAP for large-scale co-evolution analysis of protein structural interlogues. They investigated the degree of co-evolution for more than 900 family pairs in a global protein structure interactome map. They have constructed PSIMAP by systematic extraction of all protein domain contacts in the web based Protein Data Bank. Their PSIMAP contained 37387 interacting domain pairs with five or more contacts within 5 A. They have first confirmed that correlated evolution is observed extensively throughout the interacting pairs of structural families in PDB, indicating that the observation is a general property of protein evolution. The overall average correlation was 0.73 for a relatively reliable set of 454 family pairs, of which 78% showed significant correlation at 99% confidence. In total, 918 family pairs have been investigated and the correlation was 0.61 on average. But the statistical validity was weak for the family pairs with small N (the number of member domain pairs) of their research**.** This is the first step in protein classification technique two combine two properties of proteins, namely, structure comparison and interactivity.

Mr. Jong Park and Dan Bolser established a bioinformatics research group in UK named MRC-DUNN. They stated their research on protein network. They worked on structure of proteins. They also used PSIMAP concept. But the limitation is that they only focused on protein intractability and taxonomic diversity. As a result their concept did not help that much on protein structure analysis using PSIMAP concept.

Again in February 2003, Mr. Jong Park and Dan Bolser tried to integrate Biological network evolution hypothesis to protein structural interactome. PSI-MAP was used to identify all the structurally observed interactions at the structure family level. To assess the functional and evolutionary differences between the most interactive and the least interactive folds, they used the latest HIINFOLD and LOINFOLD comparison sets (Park and Bolser, 2001): high interaction structure families and low interaction structure families. The major problem of their system is that they said that scale free topology is robust. But in practical it's not true.

BMC bioinformatics research group has developed a concept of Visualization and graph-theoretic analysis of a large-scale protein structural interactome. They presented a global analysis of PSIMAP using several distinct network measures relating to centrality, interactivity, fault-tolerance, and taxonomic diversity. But to get proper structure and layout they put several proteins according to maximum similarity. As a result some proteins are placed in wrong places. And lots of scale free proteins do not get proper places. Sungsam Gong, Giseok Yoon, Insoo Jang, Dan Bolser, Panos Dafas and some other famous scientist developed PSIBase for Protein Structural Interactome map (PSIMAP). They introduced PSIbase: the PSIMAP web server and database. It contains (1) domain–domain and protein–protein interaction information from proteins whose 3D-structures are identified, (2) a protein interaction map and its viewer at protein super family and family levels, (3) protein interaction interface viewers and (4) structural domain prediction tools for possible interactions by detecting homologous matches in the Protein Data Bank (PDB) from query sequences. They developed an algorithm. According to that algorithm the basic mechanism to check interactions between any two domains or proteins is the calculation of the Euclidean distance in order to see if they are within a certain distance threshold. PSIMAP checks every possible pair of structural domains in a protein to see if there are at least five residue contacts within a 5Å distance **[18].**

Daeui Park, Semin Lee, Dan Bolser, Michael Schroeder some other scientists at beginning of 2005 have developed Comparative interactomics analysis of protein family interaction networks using PSIMAP (protein structural interactome map) They have confirmed that all the predicted protein family interactomes (the full set of protein family interactions within a proteome) of 146 species are scale-free networks, and they share a small core network comprising 36 protein families related to indispensable cellular functions. To construct the protein family interaction network in a particular proteome, they first assigned the known 3D structural families (on which PSIMAP is based) to the protein sequences. 146 completely sequenced species from the European Bioinformatics Institute (EBI) and their 578,625 protein sequences were used **(Pruess, et al., 2003).**

The above study clearly shows that yet now there is no technique has developed to classify proteins incorporating all six major properties. Though in protein grouping technique PSIMAP is one of the remarkable achievements in this context but it has some drawbacks **[1, 19]** especially in grouping the proteins in different classes based on some essential features. To get the optimum output using PSIMAP in this context researchers have to put some proteins in comparative places **[1]**. As a result actual classification cannot be done using PSIMAP. This affects bad lay out for 3-d structure design of protein **[1]**. These proteins which cannot be placed in proper groups may be termed as scale free proteins **[1, 3, 4, and 5]**. We have tried to develop a smart algorithm to put right proteins in right places with an optimum output.

Analyzing the limitations of PSIMAP our proposed algorithm has incorporated all six major properties of proteins and succeeded to eliminate any scale free protein.

## 2. LIMITATIONS OF EXISTING ALGORITHMS IN PROTEIN GROUPING

We have studied and analyzed PSIMAP (Protein Structural Interactome Map) **[1],** Visualization and graph-theoretic analysis of a large-scale protein structural interactome **[1, 9-16]** to predict some protein functions. The predicted proteins' functions are domain-domain interaction, scale free property, age and taxonomic diversity, connectivity, interaction matrix and cluster index **[1, 17]** .We gave our main attention on one of the recent functions, scale free property of proteins. According to scale free property, some proteins can not be placed any where in the whole proteins network. We have developed our algorithm based on above proteins' functions, probability theorem and graph theory to remove scale free proteins from proteins network and finally we have grouped them.

With a view to designing a special algorithm for classification of proteins, we have examined the available searching algorithm and their effectiveness for our specific purpose. It may be mentioned that as we have planned to design a tree structure for providing a good lay out for protein groups, we have given special attention to searching algorithm in analyzing the algorithms we have considered time complexity, and their applicability in our specific context. The following searching algorithms have revealed their inefficiency to fulfill our objectives:

1. Hash Table, Selection Search and Linear Search algorithms incorporating with sorting algorithm are used to search a particular key value. We have not considered these searching algorithms for our specific purpose. Although these three algorithms work efficiently on considerably small size of data **[8, 20]**. But our objective is to design an algorithm which can efficiently work on a huge database like Protein Data Bank on the Web. In fact PDB contains huge data on protein and perhaps it is the largest web based protein database **[1, 21]**.
2. Again we also have not considered A* search algorithm for our searching technique. Because A* search algorithm is used to search a shortest path from root to a given goal node **[8, 20]**. But in this field of work we do not have any goal node where the newly coming node will be placed. Rather we have to find the exact position of the newly coming protein out by dynamically.
3. The DFS and BSF algorithms are widely used for finding out shortest path from source to destination. However, as in grouping proteins as our attempt is to generate a tree rather than a graph we have discarded these algorithms too. Besides, in discarding these algorithms we have also considered their time complexities in order of 0(n+e) [8] which are very high for our objective.
4. Best-first search is the updated version of depth first search algorithm. So it also inherits properties from DFS. So for the similar reasons we have not considered this algorithm..
5. Finally Binary search tree algorithm can be considered for its less time complexity, effectiveness and efficiency **[8]**. However as in binary search tree, each node can have at most two children node which would not be adopted for our protein classification algorithm because each group of proteins have many members and all of them may have more than two children coming out from a particular node.

Considering limitations of the above stated popular search algorithms we have considered to derive a special algorithm to fulfill our specific objective. For this, we have used weighted search concept for searching and selecting the exact position of a newly coming proteins in the big protein database. We have used partially BFS concept and also DFS concept based on weighted search concept to get the desired position of the protein.

## 3. METHODOLOGY

We have designed the algorithm using incorporating six major properties of protein. We have calculated probability of each protein newly entered in the system against of existing proteins in the system. In our approach we have considered six functions for calculating probabilities based on six properties of proteins. The individual function generates probabilistic value for deriving its respective weight against that particular property. All probabilistic values generated by six individual functions will be added together. The aggregated probabilistic value will be compared with the probabilistic value of the protein that resides at the root. If there is not root protein (i.e. at the initial state), the first protein will be considered as the root and depending on the probabilistic value it can change its relative position.

Based on guided search algorithm we chose the node which has the highest probability of level 1. Then it will start calculation and comparison the probabilistic values of level 2 of selected node from level 1. Then we chose the node having highest probability and continued until getting the exact position of newly entered protein.

In this way, a super kingdom tree for all proteins will be generated.

### 3.1. DETERMINING THE BOND FACTOR

We have applied the general probability function to calculate similarity factor of proteins of each function individually

Let, if an event is A, then the probability formula for calculation probability of A is

P (A) = Total Output / Expected Output Now if there are n events, then

The total Bond Factor of all events is P (Total) = P (A1) + P (A2) + P (A3) + P (A4) + ……………. + P (An)

Using the above formulae, the similarity factor of a protein **p1** against another protein **p2** is of above functions are given below:

P (**p1.p2**.Structure) = Similarity between **p1** and **p2** with respect to structure / expected similarity of **p1** and **p2** with respect to structure

P (**p1.p2**.Sequence) = Similarity between **p1** and **p2** with respect to Sequence/ expected similarity of **p1** and **p2** with respect to Sequence

P (**p1.p2**.Connectivity) = Similarity between **p1** and **p2** with respect to Connectivity/ expected similarity of **p1** and **p2** with respect to Connectivity

P (**p1.p2**.Cluster index) = Similarity between **p1** and **p2** with respect to Cluster index / expected similarity of **p1** and **p2** with respect to Cluster index

P (**p1.p2** .Interactivity) = Similarity between **p1** and **p2** with respect to Interactivity / expected similarity of **p1** and **p2** with respect to Interactivity

P (**p1.p2**.Taxonomic and age diversity) = Similarity between **p1** and **p2** with respect to Taxonomic and age diversity / expected similarity of **p1** and **p2** with respect to Taxonomic and age diversity

So the total probability of **p1** with respect to **p2**

P (**p1.p2**) = P (**p1.p2**.Structure) + P (**p1.p2**.Sequence) + P (**p1.p2**.Connectivity) + P (**p1.p2**.Cluster index) + P (**p1.p2**.Interactivity) + P (**p1.p2**.Taxonomic and age diversity)

**A Proof of our algorithm**

To prove the efficiency of our algorithm, we have used some dummy data containing probabilistic values for each function.

Let **p1, p2, p3, p4, p5, p6, p7, p8, p9, p10, p11, p12, p13, p14, p15** are some proteins of which structure, sequence, interactivity, cluster index **[1, 17]**, connectivity and taxonomic and age diversity values known. Based on these dummy values we have proved our proposed algorithm.

***Table 1: Probabilistic values for Structure similarities of the above proteins***

| | P1 | P2 | P3 | P4 | P5 | P6 | P7 | P8 | P9 | P10 | P11 | P12 | P13 | P14 | P15 |
|---|---|---|---|---|---|---|---|---|---|---|---|---|---|---|---|
| **P1** | 100 | 40 | 20 | 80 | 35 | 28 | 60 | 35 | 70 | 10 | 05 | 30 | 10 | 10 | 95 |
| **P2** | 40 | 100 | 40 | 20 | 90 | 10 | 05 | 10 | 50 | 20 | 70 | 45 | 35 | 25 | 45 |
| **P3** | 20 | 40 | 100 | 30 | 20 | 60 | 10 | 32 | 12 | 30 | 50 | 10 | 05 | 12 | 03 |
| **P4** | 80 | 20 | 30 | 100 | 10 | 21 | 35 | 40 | 50 | 10 | 60 | 12 | 60 | 05 | 50 |
| **P5** | 35 | 90 | 20 | 10 | 100 | 00 | 30 | 20 | 60 | 12 | 35 | 73 | 13 | 40 | 10 |
| **P6** | 28 | 10 | 60 | 21 | 00 | 100 | 10 | 00 | 01 | 60 | 34 | 21 | 90 | 07 | 95 |
| **P7** | 60 | 05 | 10 | 35 | 30 | 10 | 100 | 21 | 32 | 41 | 55 | 00 | 30 | 05 | 01 |
| **P8** | 35 | 10 | 32 | 40 | 20 | 00 | 21 | 100 | 00 | 00 | 01 | 55 | 11 | 32 | 50 |
| **P9** | 70 | 50 | 12 | 50 | 60 | 01 | 32 | 00 | 100 | 90 | 12 | 35 | 21 | 24 | 90 |

| P10 | P11 | P12 | P13 | P14 | P15 |
|---|---|---|---|---|---|
| 10 | 05 | 30 | 10 | 10 | 95 |
| 20 | 70 | 45 | 35 | 25 | 45 |
| 30 | 50 | 10 | 05 | 12 | 03 |
| 10 | 60 | 12 | 60 | 05 | 50 |
| 12 | 35 | 73 | 13 | 40 | 10 |
| 60 | 34 | 21 | 90 | 07 | 95 |
| 41 | 55 | 00 | 30 | 05 | 01 |
| 00 | 01 | 55 | 11 | 32 | 50 |
| 90 | 12 | 35 | 21 | 24 | 90 |
| 100 | 00 | 00 | 00 | 10 | 75 |
| 00 | 100 | 30 | 90 | 23 | 56 |
| 00 | 30 | 100 | 55 | 21 | 35 |
| 00 | 90 | 55 | 100 | 23 | 01 |
| 10 | 23 | 21 | 23 | 100 | 30 |
| 75 | 56 | 35 | 01 | 30 | 100 |

***Table 2: Sequence similarities of the above proteins***

| | **P1** | **P2** | **P3** | **P4** | **P5** | **P6** | **P7** | **P8** | **P9** | **P10** | **P11** | **P1** | **P13** | **P14** | **P15** |
|---|---|---|---|---|---|---|---|---|---|---|---|---|---|---|---|
| **P1** | 100 | 45 | 10 | 70 | 45 | 20 | 50 | 30 | 75 | 15 | 15 | 40 | 20 | 13 | 85 |
| **P2** | 45 | 100 | 30 | 30 | 95 | 05 | 15 | 11 | 45 | 15 | 60 | 50 | 30 | 30 | 40 |
| **P3** | 10 | 30 | 100 | 30 | 20 | 60 | 10 | 32 | 12 | 30 | 50 | 10 | 05 | 12 | 03 |
| **P4** | 70 | 30 | 30 | 100 | 05 | 20 | 30 | 50 | 40 | 15 | 65 | 10 | 63 | 07 | 48 |
| **P5** | 45 | 95 | 20 | 05 | 100 | 00 | 33 | 25 | 61 | 15 | 30 | 70 | 15 | 45 | 12 |
| **P6** | 20 | 05 | 60 | 20 | 00 | 100 | 10 | 00 | 01 | 60 | 34 | 21 | 90 | 07 | 95 |
| **P7** | 50 | 15 | 10 | 30 | 33 | 10 | 100 | 21 | 32 | 41 | 55 | 00 | 30 | 05 | 01 |
| **P8** | 30 | 11 | 32 | 50 | 25 | 00 | 21 | 100 | 01 | 10 | 21 | 45 | 15 | 35 | 53 |
| **P9** | 75 | 45 | 12 | 40 | 61 | 01 | 32 | 01 | 100 | 90 | 12 | 35 | 21 | 24 | 90 |
| **P1** | 15 | 15 | 30 | 15 | 15 | 60 | 41 | 10 | 90 | 100 | 00 | 00 | 00 | 10 | 75 |
| **P1** | 15 | 60 | 50 | 65 | 30 | 34 | 55 | 21 | 12 | 00 | 100 | 30 | 90 | 23 | 56 |
| **P1** | 40 | 50 | 10 | 10 | 70 | 21 | 00 | 45 | 35 | 00 | 30 | 10 | 55 | 21 | 35 |
| **P1** | 20 | 30 | 05 | 63 | 15 | 90 | 30 | 15 | 21 | 00 | 90 | 55 | 100 | 23 | 01 |
| **P1** | 13 | 30 | 12 | 07 | 45 | 07 | 05 | 35 | 24 | 10 | 23 | 21 | 23 | 100 | 30 |
| **P1** | 85 | 40 | 03 | 48 | 12 | 95 | 01 | 53 | 90 | 75 | 56 | 35 | 01 | 30 | 100 |

***Table 3: Interactivity similarities of the above proteins***

| | **P1** | **P2** | **P3** | **P4** | **P5** | **P6** | **P7** | **P8** | **P9** | **P10** | **P11** | **P12** | **P13** | **P14** | **P15** |
|---|---|---|---|---|---|---|---|---|---|---|---|---|---|---|---|
| **P1** | 100 | 48 | 10 | 73 | 48 | 17 | 50 | 29 | 75 | 12 | 12 | 40 | 17 | 13 | 85 |
| **P2** | 48 | 100 | 29 | 29 | 95 | 05 | 12 | 11 | 48 | 12 | 60 | 50 | 29 | 29 | 40 |
| **P3** | 10 | 29 | 100 | 29 | 17 | 60 | 10 | 32 | 12 | 29 | 50 | 10 | 05 | 12 | 03 |
| **P4** | 73 | 29 | 29 | 100 | 05 | 17 | 29 | 50 | 40 | 12 | 62 | 10 | 63 | 07 | 48 |
| **P5** | 48 | 95 | 17 | 05 | 100 | 00 | 33 | 25 | 61 | 12 | 29 | 73 | 12 | 48 | 12 |
| **P6** | 17 | 05 | 60 | 17 | 00 | 100 | 10 | 00 | 01 | 60 | 34 | 21 | 90 | 07 | 95 |
| **P7** | 50 | 12 | 10 | 29 | 33 | 10 | 100 | 21 | 32 | 41 | 55 | 00 | 29 | 05 | 01 |
| **P8** | 29 | 11 | 32 | 50 | 25 | 00 | 21 | 100 | 05 | 12 | 17 | 40 | 10 | 32 | 50 |
| **P9** | 75 | 48 | 12 | 40 | 61 | 01 | 32 | 05 | 100 | 90 | 12 | 35 | 21 | 24 | 90 |
| **P10** | 12 | 12 | 29 | 12 | 12 | 60 | 41 | 12 | 90 | 100 | 00 | 00 | 00 | 10 | 75 |
| **P11** | 12 | 60 | 50 | 62 | 29 | 34 | 55 | 17 | 12 | 00 | 100 | 29 | 90 | 23 | 56 |
| **P12** | 40 | 50 | 10 | 10 | 73 | 21 | 00 | 40 | 35 | 00 | 29 | 100 | 55 | 21 | 35 |
| **P13** | 17 | 29 | 05 | 63 | 12 | 90 | 29 | 10 | 21 | 00 | 90 | 55 | 100 | 23 | 01 |
| **P14** | 13 | 29 | 12 | 07 | 48 | 07 | 05 | 32 | 24 | 10 | 23 | 21 | 23 | 100 | 29 |
| **P15** | 85 | 40 | 03 | 48 | 12 | 95 | 01 | 50 | 90 | 75 | 56 | 35 | 01 | 29 | 100 |

***Table 4: Connectivity similarities of the above proteins***

| | **P1** | **P2** | **P3** | **P4** | **P5** |
|---|---|---|---|---|---|
| **P1** | 100 | 42 | 10 | 72 | 42 |
| **P2** | 42 | 100 | 32 | 32 | 95 |
| **P3** | 10 | 32 | 100 | 32 | 23 |
| **P4** | 72 | 32 | 32 | 100 | 05 |
| **P5** | 42 | 95 | 23 | 05 | 100 |
| **P6** | 23 | 05 | 60 | 23 | 00 |
| **P7** | 50 | 15 | 10 | 32 | 33 |
| **P8** | 32 | 11 | 32 | 50 | 25 |
| **P9** | 75 | 42 | 12 | 35 | 61 |
| **P10** | 15 | 15 | 32 | 15 | 15 |
| **P11** | 15 | 60 | 50 | 65 | 32 |
| **P12** | 35 | 50 | 10 | 10 | 72 |
| **P13** | 23 | 32 | 05 | 63 | 15 |
| **P14** | 13 | 32 | 12 | 07 | 42 |
| **P15** | 85 | 35 | 03 | 48 | 12 |

| | | | | | | | | | | | | | | | |
|---|---|---|---|---|---|---|---|---|---|---|---|---|---|---|---|
| **P6** | 23 | 05 | 60 | 23 | 00 | 100 | 10 | 00 | 01 | 60 | 34 | 21 | 92 | 07 | 95 |
| **P7** | 50 | 15 | 10 | 32 | 33 | 10 | 100 | 21 | 32 | 41 | 55 | 00 | 32 | 05 | 01 |
| **P8** | 32 | 11 | 32 | 50 | 25 | 00 | 21 | 100 | 01 | 10 | 21 | 42 | 15 | 35 | 53 |
| **P9** | 75 | 42 | 12 | 35 | 61 | 01 | 32 | 01 | 100 | 92 | 12 | 35 | 21 | 24 | 92 |
| **P10** | 15 | 15 | 32 | 15 | 15 | 60 | 41 | 10 | 92 | 100 | 00 | 00 | 00 | 10 | 75 |
| **P11** | 15 | 60 | 50 | 65 | 32 | 34 | 55 | 21 | 12 | 00 | 100 | 32 | 92 | 23 | 56 |
| **P12** | 35 | 50 | 10 | 10 | 72 | 21 | 00 | 42 | 35 | 00 | 32 | 100 | 55 | 21 | 35 |
| **P13** | 23 | 32 | 05 | 63 | 15 | 92 | 32 | 15 | 21 | 00 | 92 | 55 | 100 | 23 | 01 |
| **P14** | 13 | 32 | 12 | 07 | 42 | 07 | 05 | 35 | 24 | 10 | 23 | 21 | 23 | 100 | 32 |
| **P15** | 85 | 35 | 03 | 48 | 12 | 95 | 01 | 53 | 92 | 75 | 56 | 35 | 01 | 32 | 100 |

***Table 5: Cluster index similarities of the above proteins***

| | **P1** | **P2** | **P3** | **P4** | **P5** | **P6** | **P7** | **P8** | **P9** | **P10** | **P11** | **P12** | **P13** | **P14** | **P15** |
|---|---|---|---|---|---|---|---|---|---|---|---|---|---|---|---|
| **P1** | 100 | 45 | 10 | 65 | 45 | 20 | 50 | 30 | 75 | 12 | 12 | 38 | 20 | 13 | 85 |
| **P2** | 45 | 100 | 30 | 30 | 95 | 05 | 12 | 11 | 45 | 12 | 60 | 50 | 30 | 30 | 38 |
| **P3** | 10 | 30 | 100 | 30 | 20 | 60 | 10 | 32 | 12 | 30 | 50 | 10 | 05 | 12 | 03 |
| **P4** | 65 | 30 | 30 | 100 | 05 | 20 | 30 | 50 | 38 | 12 | 65 | 10 | 63 | 07 | 48 |
| **P5** | 45 | 95 | 20 | 05 | 100 | 00 | 33 | 25 | 61 | 12 | 30 | 65 | 12 | 45 | 12 |
| **P6** | 20 | 05 | 60 | 20 | 00 | 100 | 10 | 00 | 01 | 60 | 34 | 21 | 90 | 07 | 95 |
| **P7** | 50 | 12 | 10 | 30 | 33 | 10 | 100 | 21 | 32 | 41 | 55 | 00 | 30 | 05 | 01 |
| **P8** | 30 | 11 | 32 | 50 | 25 | 00 | 21 | 100 | 01 | 10 | 21 | 45 | 12 | 28 | 53 |
| **P9** | 75 | 45 | 12 | 38 | 61 | 01 | 32 | 01 | 100 | 90 | 12 | 28 | 21 | 24 | 90 |
| **P10** | 12 | 12 | 30 | 12 | 12 | 60 | 41 | 10 | 90 | 100 | 00 | 00 | 00 | 10 | 75 |
| **P11** | 12 | 60 | 50 | 65 | 30 | 34 | 55 | 21 | 12 | 00 | 100 | 30 | 90 | 23 | 56 |
| **P12** | 38 | 50 | 10 | 10 | 65 | 21 | 00 | 45 | 28 | 00 | 30 | 100 | 55 | 21 | 28 |
| **P13** | 20 | 30 | 05 | 63 | 12 | 90 | 30 | 12 | 21 | 00 | 90 | 55 | 100 | 23 | 01 |
| **P14** | 13 | 30 | 12 | 07 | 45 | 07 | 05 | 28 | 24 | 10 | 23 | 21 | 23 | 100 | 30 |
| **P15** | 85 | 38 | 03 | 48 | 12 | 95 | 01 | 53 | 90 | 75 | 56 | 28 | 01 | 30 | 100 |

***Table 6: Taxonomic and age diversity similarities of the above proteins***

| | **P1** | **P2** | **P3** | **P4** | **P5** | **P6** | **P7** | **P8** | **P9** | **P10** | **P11** | **P12** | **P13** | **P14** | **P15** |
|---|---|---|---|---|---|---|---|---|---|---|---|---|---|---|---|
| **P1** | 100 | 45 | 10 | 70 | 45 | 25 | 48 | 32 | 75 | 15 | 15 | 40 | 25 | 13 | 85 |
| **P2** | 45 | 100 | 32 | 32 | 95 | 05 | 15 | 11 | 45 | 15 | 60 | 48 | 32 | 32 | 40 |
| **P3** | 10 | 32 | 100 | 32 | 25 | 60 | 10 | 32 | 12 | 32 | 48 | 10 | 05 | 12 | 03 |
| **P4** | 70 | 32 | 32 | 100 | 05 | 25 | 32 | 48 | 40 | 15 | 65 | 10 | 63 | 07 | 48 |
| **P5** | 45 | 95 | 25 | 05 | 100 | 02 | 33 | 25 | 61 | 15 | 32 | 70 | 15 | 45 | 12 |
| **P6** | 25 | 05 | 60 | 25 | 02 | 100 | 10 | 02 | 01 | 60 | 34 | 21 | 90 | 07 | 95 |
| **P7** | 48 | 15 | 10 | 32 | 33 | 10 | 100 | 21 | 32 | 41 | 55 | 02 | 32 | 05 | 01 |
| **P8** | 32 | 11 | 32 | 48 | 25 | 02 | 21 | 100 | 01 | 10 | 21 | 45 | 15 | 35 | 53 |
| **P9** | 75 | 45 | 12 | 40 | 61 | 01 | 32 | 01 | 100 | 90 | 12 | 35 | 21 | 24 | 90 |
| **P10** | 15 | 15 | 32 | 15 | 15 | 60 | 41 | 10 | 90 | 100 | 02 | 02 | 02 | 10 | 75 |
| **P11** | 15 | 60 | 48 | 65 | 32 | 34 | 55 | 21 | 12 | 02 | 100 | 32 | 90 | 23 | 56 |
| **P12** | 40 | 48 | 10 | 10 | 70 | 21 | 02 | 45 | 35 | 02 | 32 | 100 | 55 | 21 | 35 |
| **P13** | 25 | 32 | 05 | 63 | 15 | 90 | 32 | 15 | 21 | 02 | 90 | 55 | 100 | 23 | 01 |
| **P14** | 13 | 32 | 12 | 07 | 45 | 07 | 05 | 35 | 24 | 10 | 23 | 21 | 23 | 100 | 32 |
| **P15** | 85 | 40 | 03 | 48 | 12 | 95 | 01 | 53 | 90 | 75 | 56 | 35 | 01 | 32 | 100 |

*Table 7: Total probability of all proteins with respect to structure, sequence, connectivity, cluster index, interactivity and taxonomic and age diversity*

| | P1 | P2 | P3 | P4 | P5 | P6 | P7 | P8 | P9 | P10 | P11 | P12 | P13 | P14 | P15 |
|---|---|---|---|---|---|---|---|---|---|---|---|---|---|---|---|
| P1 | 6 | | | | | | | | | | | | | | |
| P2 | 2.65 | 6 | | | | | | | | | | | | | |
| P3 | 0.70 | 1.93 | 6 | | | | | | | | | | | | |
| P4 | 4.30 | 1.73 | 1.83 | 6 | | | | | | | | | | | |
| P5 | 2.60 | 5.65 | 1.25 | 0.35 | 6 | | | | | | | | | | |
| P6 | 1.33 | 0.35 | 3.80 | 1.26 | 0.02 | 6 | | | | | | | | | |
| P7 | 3.08 | 0.74 | 0.60 | 1.88 | 1.95 | 0.60 | 6 | | | | | | | | |
| P8 | 1.88 | 0.65 | 1.92 | 2.88 | 1.45 | 0.02 | 1.26 | 6 | | | | | | | |
| P9 | 4.45 | 2.75 | 0.72 | 2.43 | 3.65 | 0.06 | 1.92 | 0.09 | 6 | | | | | | |
| P10 | 0.79 | 0.89 | 1.83 | 0.79 | 0.81 | 3.60 | 2.46 | 0.52 | 5.42 | 6 | | | | | |
| P11 | 0.74 | 3.70 | 2.98 | 3.82 | 1.88 | 2.04 | 3.30 | 1.02 | 0.72 | 0.02 | 6 | | | | |
| P12 | 2.23 | 2.93 | 0.60 | 0.62 | 4.23 | 1.26 | 0.02 | 2.17 | 2.03 | 0.02 | 1.83 | 6 | | | |
| P13 | 1.15 | 1.88 | 0.30 | 3.75 | 0.82 | 5.42 | 1.83 | 0.78 | 1.26 | 0.02 | 5.42 | 3.30 | 6 | | |
| P14 | 0.75 | 1.78 | 0.72 | 0.40 | 2.65 | 0.42 | 0.30 | 1.97 | 1.44 | 0.60 | 1.38 | 1.26 | 1.38 | 6 | |
| P15 | 5.20 | 2.38 | 0.18 | 2.90 | 0.70 | 5.70 | 0.06 | 3.12 | 5.42 | 4.50 | 3.36 | 2.03 | 0.06 | 1.83 | 6 |

Now based on the total Bond Factor stated in Table 7, the proposed algorithm has been simulated with a view to generating a tree structure using all 15 proteins leaving no scale free protein.

Let the sequence of entering proteins are **p1, p2, p3, p4, p5, p6, p7, p8, p9, p10, p11, p12, p13, p14, p15.**

Now using the respective value for Bond Factor. Let the sequence of entering proteins are **p1, p2, p3, p4, p5, p6, p7, p8, p9, p10, p11, p12, p13, p14, p15.**

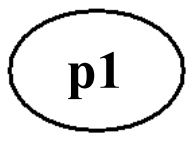

*Figure 1: Step1, Entry of p1*

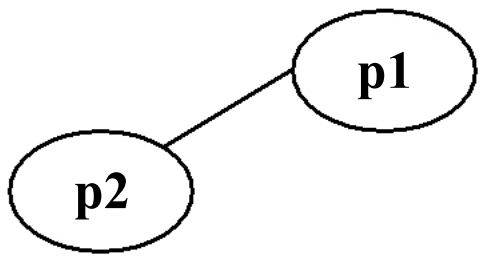

*Figure 2: Step2, Entry of p2*

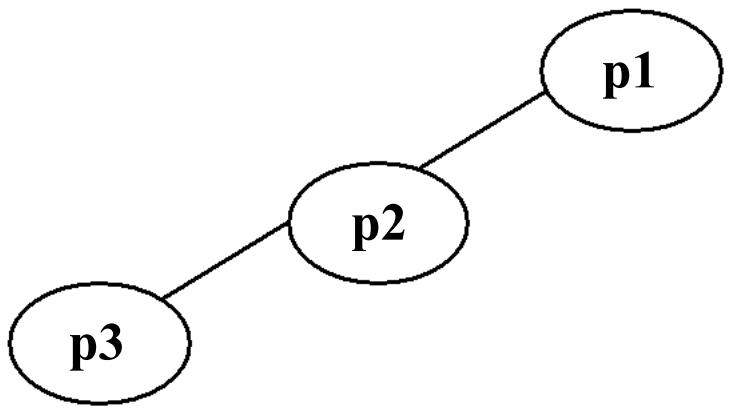

*Figure 3: Step3, Entry of p3*

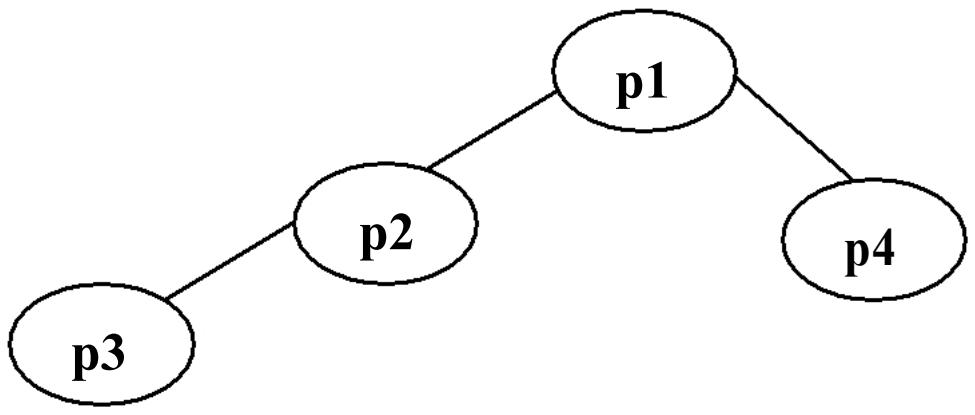

*Figure 4: Step4, Entry of p4*

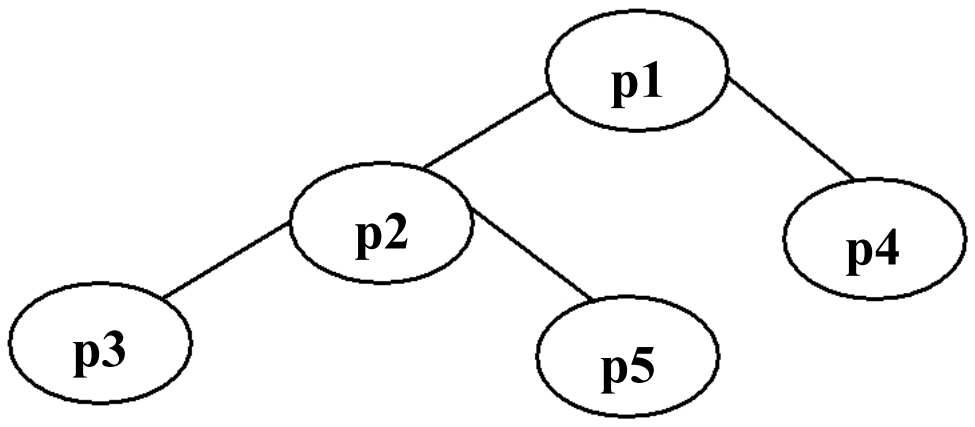

*Figure 5: Step5, Entry of p5*

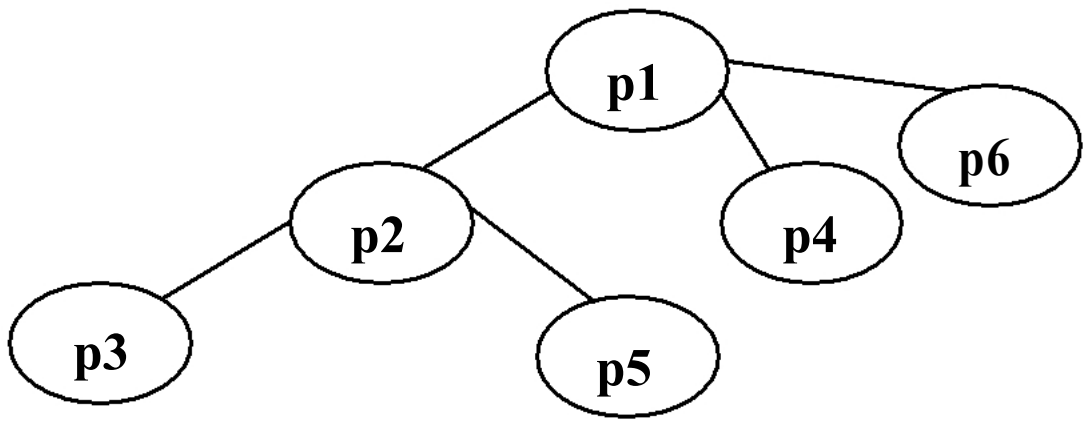

*Figure 6: Step6, Entry of p6*

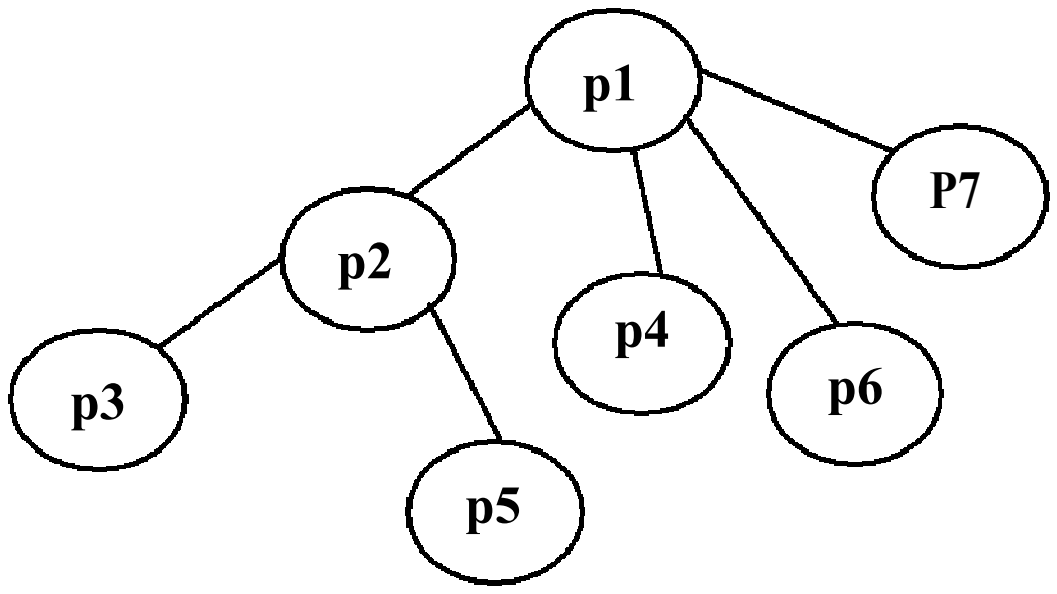

*Figure 7: Step7, Entry of p7*

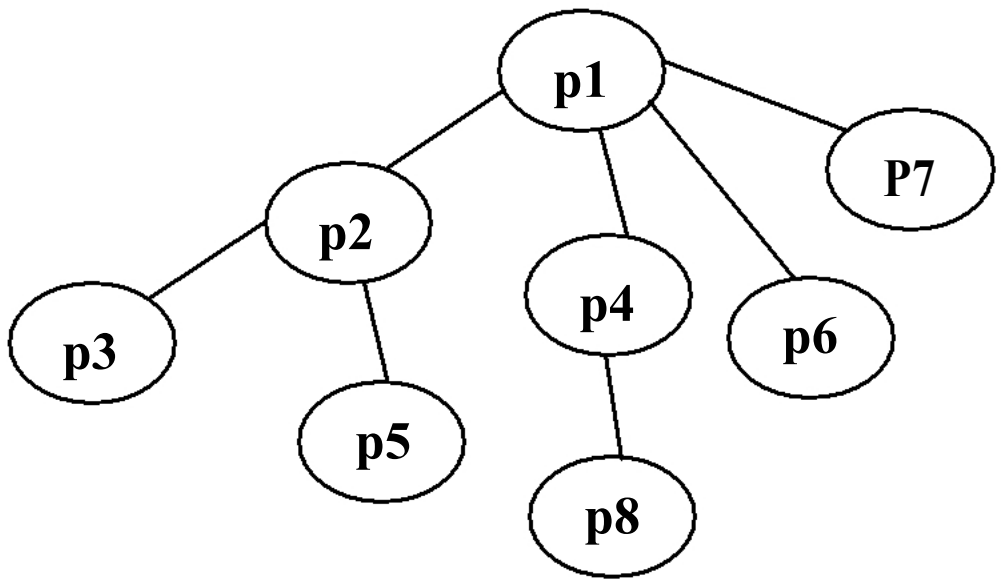

*Figure 8: Step8, Entry of p8*

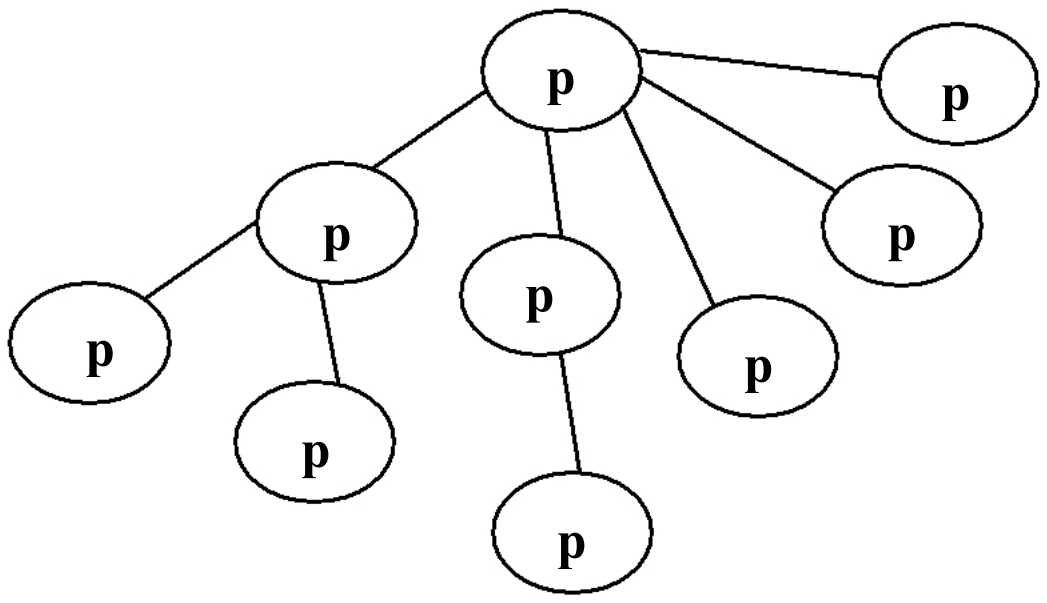

*Figure 9: Step9, Entry of p9*

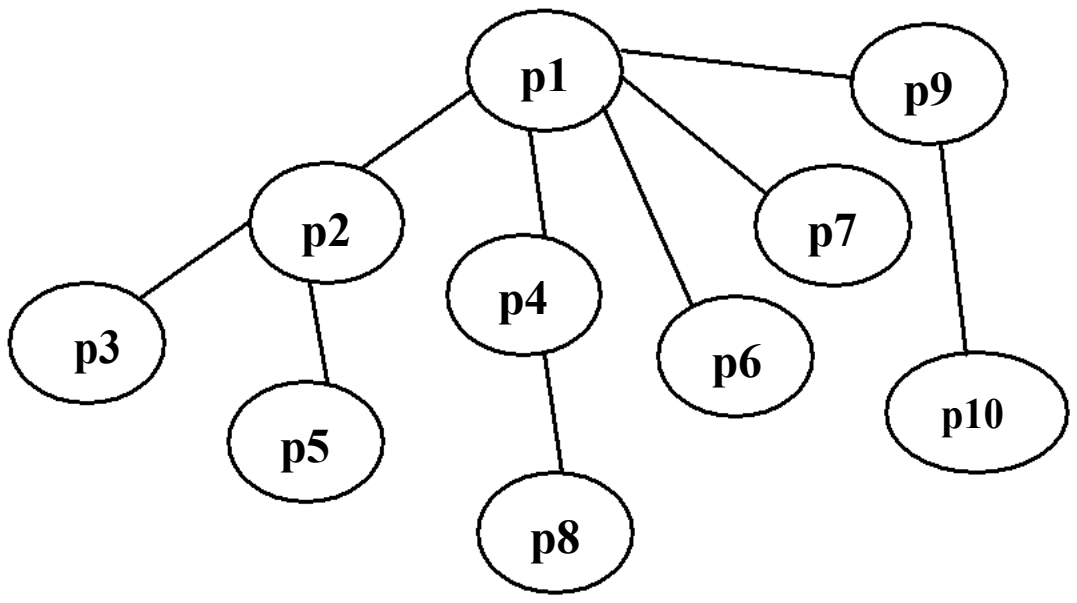

*Figure 10: Step10, Entry of p10*

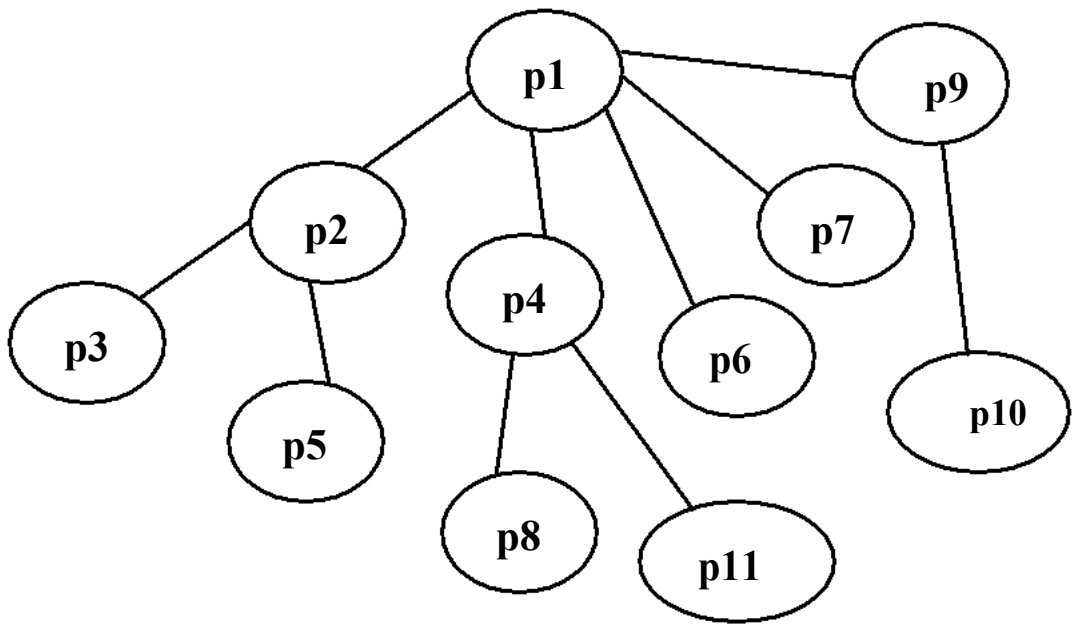

*Figure 11: Step11, Entry of p11*

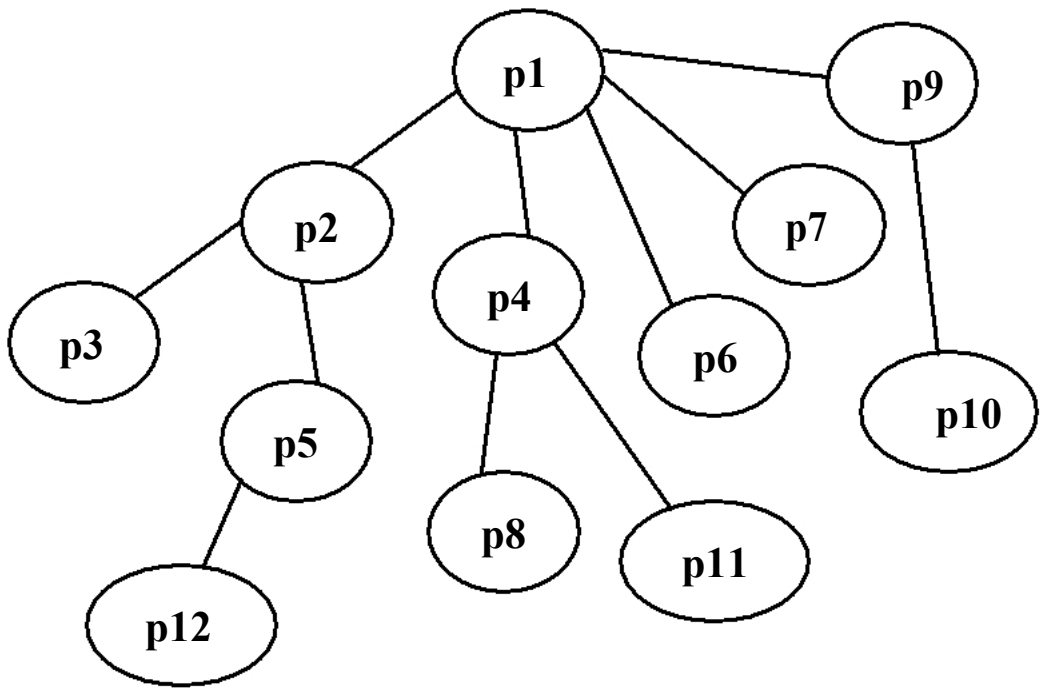

*Figure 12: Step12, Entry of p12*

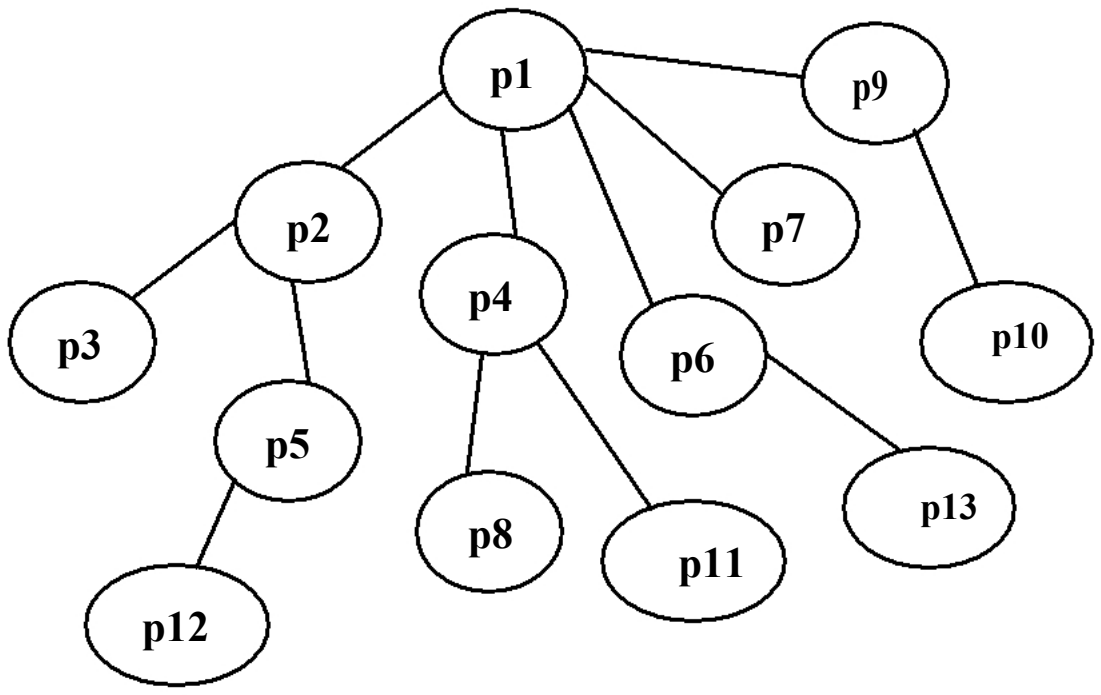

*Figure 13: Step13, Entry of p13*

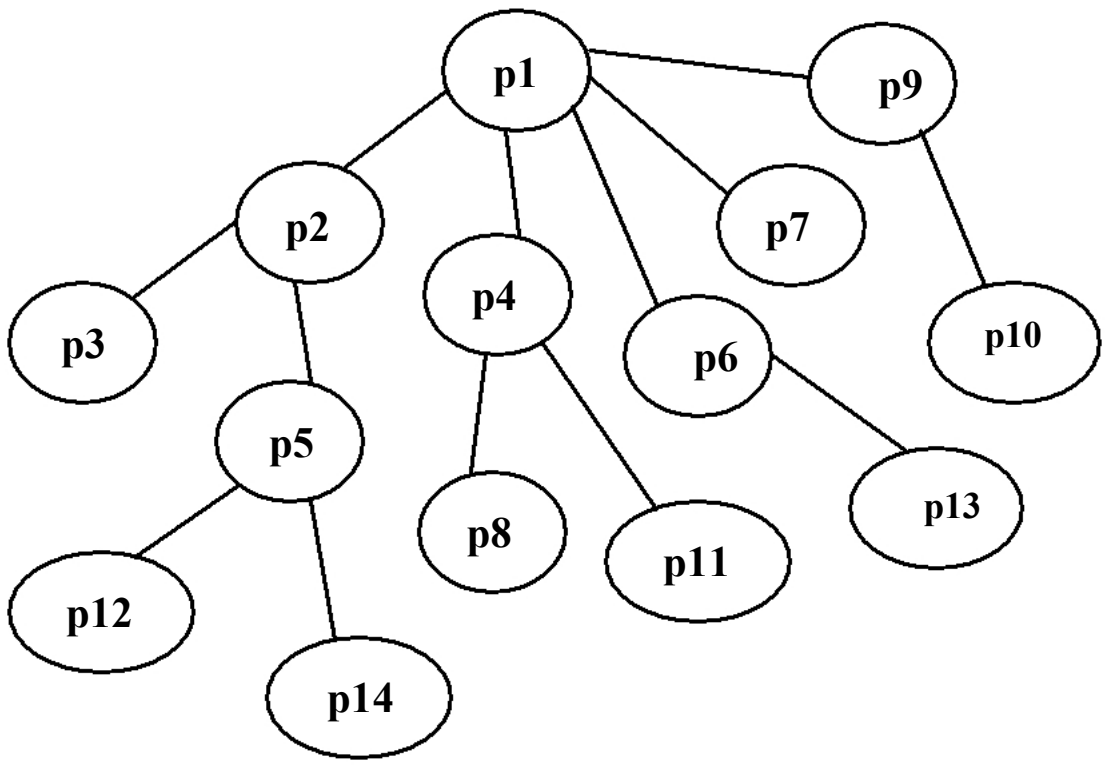

*Figure 14: Step14, Entry of p14*

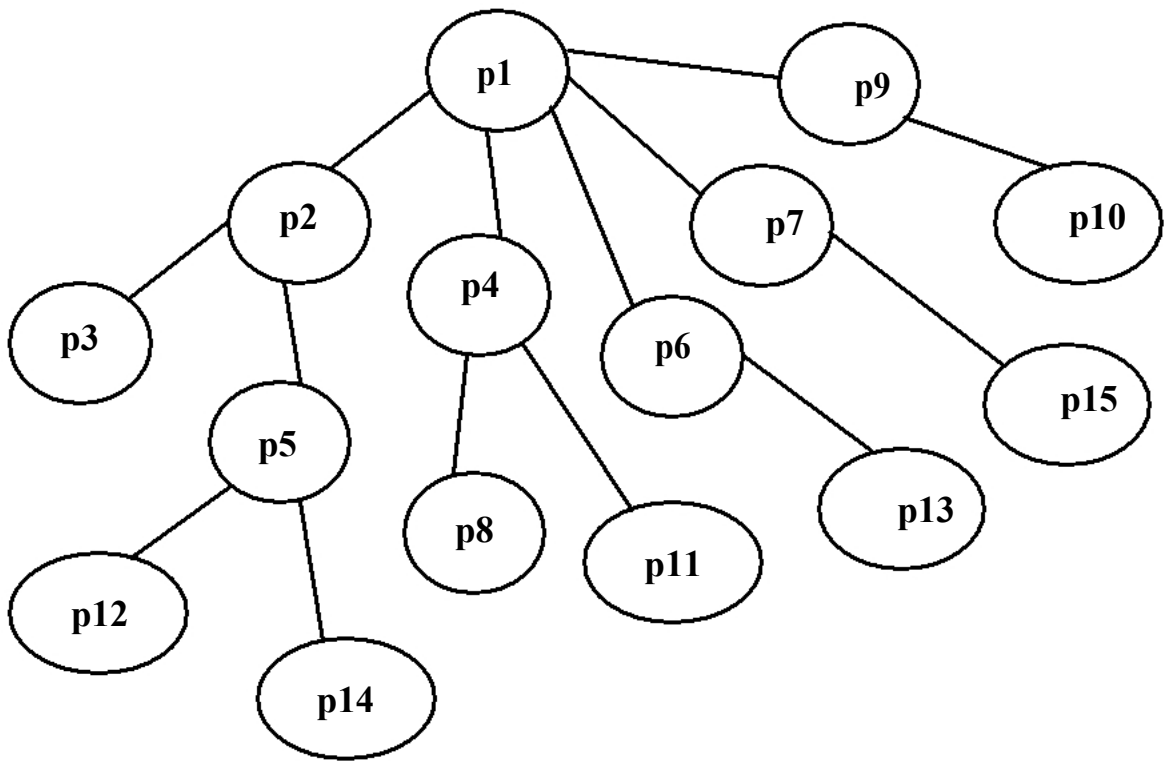

*Figure 15: Step15, Entry of p15*

### 3.2. PSEUDO CODE:

The simple pseudo code of the algorithm is given below

**Begin:**

```
Protein proteinFile1;          // declare a protein file variable
Protein proteinFile2;          // declare a protein file variable
proteinFile1= Read a protein;  // Read a protein File
Parent [0] = proteinFile1;     // Initialize the parent array by file proteinFile1 as root
TotalProbability = 0;          // initialize the total probability as zero
MaximumProbability = 0;
Structural probability;        // declare variable for structural probability
Sequential probability;        // declare variable for sequential probability
Interactivity probability;     // declare variable for interactivity probability
Cluster index probability;     // declare variable for cluster index probability
Connectivity probability;      // declare variable for connectivity probability
Taxonomic and age diversity probability; // declare variable for taxonomic
                               //and age diversity probability
TreeNode;
                               // declare TreeNode as a node of tree
CurrentSelectedNode;
                               // declare CurrentSelectedNode as a node of Tree
Do
    {
      proteinFile2 = Read a protein;
      TreeNode = Parent [0];
      While (Location is not fix)
          Do
              CurrentSelectedNode = TreeNode;
              For each node n of TreeNode
                  Do
                      Calculate TotalProbability = StructuralProbability (n, proteinFile2) + SequentialProbability (n, proteinFile2) + InteractivityProbability (n, proteinFil e2) + ClusterIndexProbability (n,proteinFile2) + ConnectivityProbability (n, proteinFile2)+ TaxonomicAgeDiversityProbability (n, proteinfile2);

// Use different functions to calculate the total

// probability

                      If TotalProbability > = MaximumProbability
                      Then
                          MaximumProbability = TotalProbability;
                          CurrentSelectedNode = n;
                      // End If
              // End For
              If (all nodes of TreeNode are finished and TreeNode = CurrentSelectedNode)

              Then
                  TreeNode -> Child = proteinFile2;  // put the position of the protein which
                                                     was newly read
                  Break; // out from inner while loop
              Else
                  TreeNode = CurrentSelectedNode; // select next parent node
              // End If
      // End While
    }
While (! End of proteins)
```

40 // **End While**

**End;**

### 3.3. TIME COMPLEXITY OF THE PROPOSED ALGORITHM

We considered only time complexity. The T (A) is total time of compilation and execution by the algorithm. The compile time doesn't depend on the instance characteristics. So we just concern ourselves with the run time of the algorithm.

The time complexity of the proposed algorithm
Worst case: T (A) = O (n) where n= number of protein file or node
Best case: T (A) = O (l) where l = level of the tree

## 4. CONCLUSION

Our algorithm for protein classification has incorporated the major six properties of protein, namely, a) Structure comparison b)Sequence Comparison c) Connectivity d) Cluster Index e) Interactivity f) Taxonomic and age diversity. Integration of all properties in a single protein group technique provides a new dimension in protein grouping. Unlike PSIMAP technique this will leave any scale free protein that to be created using this algorithm. The simulation of the algorithm using dummy data has been proved our assertion. Moreover, in term of time complexity if we consider huge protein database then it will be more efficient comparing with other existing protein grouping techniques.

However, the success of this algorithm depends on the functions that are to be used to generate probabilistic value for each protein in the proposed algorithm. But our study has revealed that some of such functions based on the properties of proteins are yet to be derived in different bioinformatics research lab **[7]** such as cluster index **[1, 17]**, connectivity and interactivity. If the respective functions for cluster index, connectivity and interactivity are achieved then our algorithm will be the protein grouping technique.